# Structural Organization of Space Polymers

Julie E. M. McGeoch[1]* and Malcolm W. McGeoch[2]

[1] Department of Molecular and Cellular Biology, Harvard University, 52 Oxford St., Cambridge MA  02138, USA.
*Corresponding author. E-mail: mcgeoch@fas.harvard.edu
[2] PLEX Corporation, 275 Martine St., Suite 100, Fall River, MA 02723, USA.

**Abstract**
Extra-terrestrial polymers of glycine with iron have been characterized by mass spectrometry to have a core mass of 1494Da with dominant rod-like variants at m/z 1567 and m/z 1639 [1]. Several principal macro-structural morphologies are observed in solvent extracts from CV3 class meteoritic material. The first is an extended sheet of linked triskelia containing the 1494Da core entity that encloses gas bubbles in the solvent [1]. A second is of fiber-like crystals found here, via X-ray diffraction, to be multiple-walled nanotubes made from a square lattice of the 1494Da polymer. A third is a dispersion of floating phantom-like short tubes of up to 100micron length [1] with characteristic angled bends that suggest the influence of a specific underlying protein structure. Here it is proposed that the angled tubes are the observable result of a space-filling super-polymerization of 1638Da polymer subunits guided by the tetragonal symmetry of linking silicon bonds. Distorted hexagonal sheets are linked by perpendicular subunits in a three-dimensional hexagonal diamond structure to fill the largest possible volume. This extended very low-density structure is conjectured to have dominated in a process of chemical selection because it captured a maximum amount of molecular raw material in the ultra-low density of molecular clouds or of the proto-solar nebula. This could have led ultimately to the accretion of the earliest planetary bodies.

## 1. Introduction

Since the first observation of polymers of amino acids in meteoritic material [2,3] there has been a report [4] of a 4641Da polymer in both Acfer 086 and Allende CV3 class meteorites. Subsequent higher resolution mass spectrometry [1] has revealed a core structure at 1494Da that directly assembles via silicon bonding into three-legged triskelia [1] observed at m/z 4641 and higher multiples of this. Mass spectrometry therefore suggests that the triskelia bond into extended sheets that in principle can cover any type of plane or curved surface via assembly into a tiled sheet of hexagons and pentagons. One example is the layer observed on the wall of spherical gas bubbles in liquid [1]. Fibrous crystals from the same extract are here analyzed by X-ray diffraction, to confirm many aspects of the 1494Da polymer structure and its super-assembly into macroscopic shapes.

A striking morphology to be seen at the interphase layer of the Folch extraction vial (Figure 1) is that of the angled phantom-like tubes shown in Figure 2. These appear less than 24 hours after exposure of micron-sized meteorite particles to solvent extraction by chloroform/methanol/water as described previously [1,3,4] and in S1 methods. The total volume of these is greater than that of the initial particles, and yet amino acid polymers comprise less than a few percent of the meteoritic contents. Here it is conjectured that the observed rod-like 1638Da subunits [1] are able to form a very low-density space-filling structure, with tetragonal linkage at silicon atoms. The internal space defined by this structure is filled with solvent under our experimental regime, but when "dry" the structure has an estimated density of 32mg cm$^{-3}$, presented below.

In this paper we first discuss the X-ray diffraction data (Section 2) then focus in sections 3, 4 and 5 on the structure underlying the shapes in Figure 2 before presenting discussion and conclusions in Section 6.

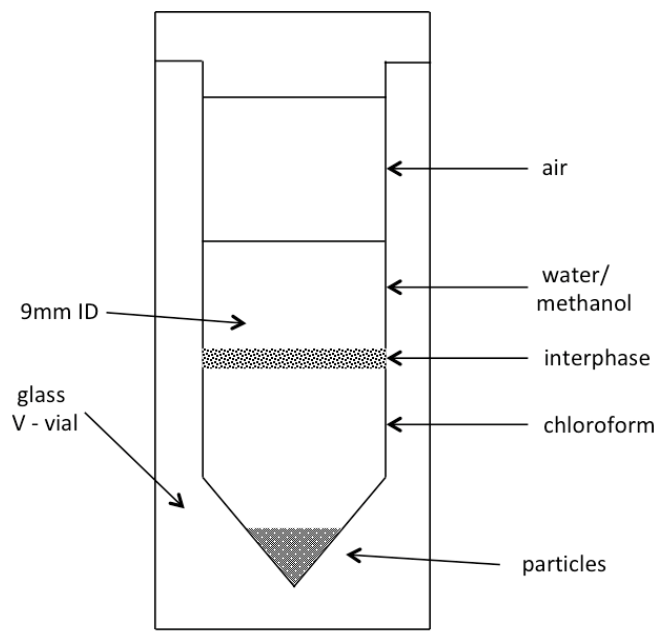

**Figure 1. Folch extraction glass V-vial. 3-D structures are at the interphase between water/methanol and chloroform. Particles sink in chloroform.**

**2. Vesicles and nano-tubules: X-ray diffraction.**
Characterized by mass spectrometry as having a core mass of 1494Da and comprising double stranded anti-parallel poly-glycine of 11 residues each side that are closed out by iron atoms [1], this rod-like protein bonds in trimers (as triskelia) to form a continuous sheet [1]. However, optical microscopy of a (disturbed) sample of the sheet material enveloping gas bubbles [1] showed a tangled cluster of fibers. Following solvent extraction in a closed glass V-vial (Figure 1), a solution of meteoritic extract (from Acfer 086) concentrated to dryness via slow evaporation over 1 month was rehydrated with slight addition of water to develop the fiber-like crystals shown in Figure 3. Three of these in a cluster were mounted on a nylon loop and subjected to X-ray diffraction at 0.979A (Angstrom) on the APS Argonne synchrotron (methods in S1).

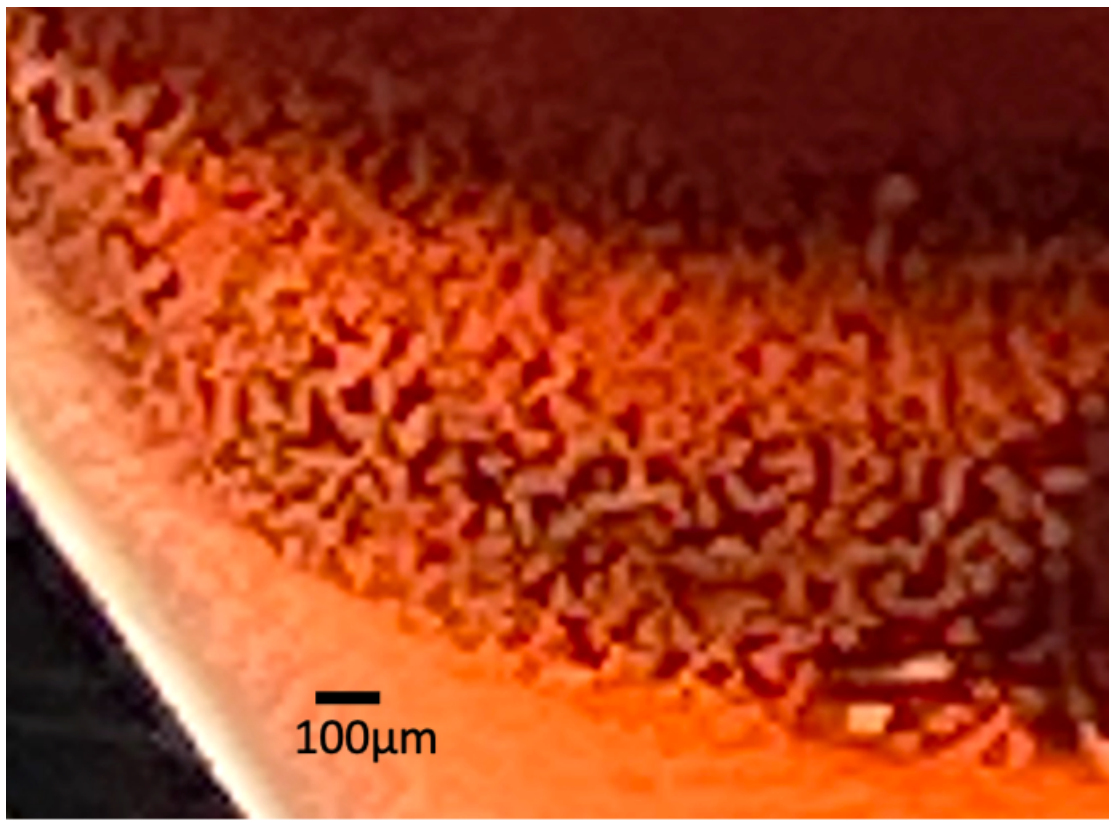

**Figure 2 Three-dimensional macrostructures from KABA at the interphase layer. Horizontal view into the layer. The red coloration is from Sudan III dye added to better visualize the 3D structures. Similar structures are seen in Acfer 086 [1].**

Figure 4 shows a diffraction frame from one of these fibers. The X-ray data frames each represent integration over a 0.2 degree oscillation. The pattern remains constant across many tens of frames indicating a circular tube structure. Regardless of orientation a constant central motif is seen, comprising short opposed arcs at nominal first order spacing of 24.19A and 16.12A, with a much less intense additional arc at 12.06A and the hint of an arc at 9.64A. Apart from this fixed central motif there were mostly-full rings at 6.31A and 5.29A, with fainter rings at 4.80A and 3.52A, the 6.31A and 5.29A rings showing four-fold symmetry and a $45^0$ orientation to the central motif.

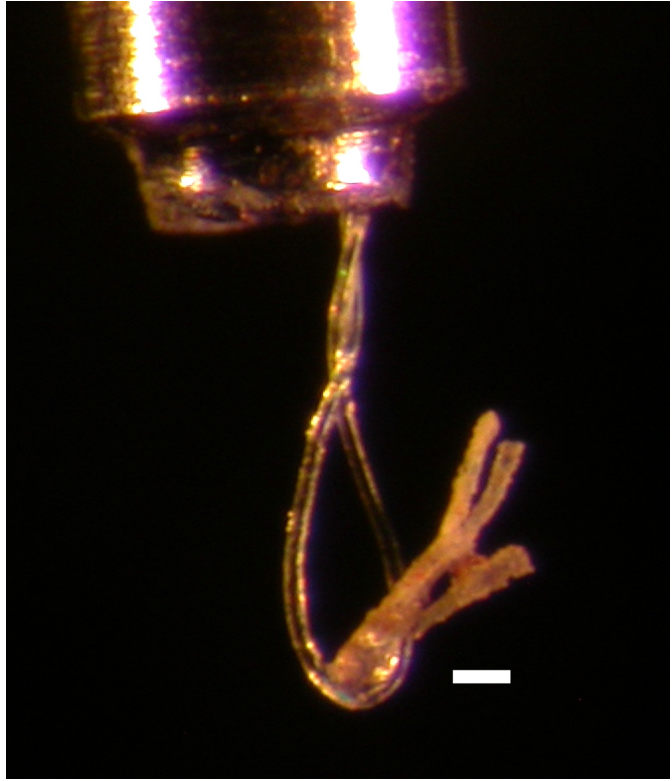

**Figure 3 Fiber-like crystals, Acfer 086, on loop (scale bar = 100μm)**

The angle independence in the central motif suggested concentric round tubes with multiple diffracting layers, as illustrated in Figure 5. In turn, this suggested that the tube walls could be gently curved lattices comprised of the rod-like entities identified in mass spectrometry [1] of mass approximately 1600Da, and typical length about 5nm. Several questions arise:

**1.** Which of the 1494, 1566, 1638 Da polymers is in the crystal?
**2.** What is the symmetry of the lattice? Possibilities to consider are:
a) hexagonal, based upon the three-fold vertex of a triskelion [1] which can be oriented in two ways relative to the axis of the tube.
b) rectangular, based upon a distortion of the triskelion vertex into a "T" pattern, in which 2:1 rectangular "bricks" can be stacked to form the mesh, again oriented in two ways parallel to the tube axis.
c) square, based upon a four-fold vertex. This presents a single orientation in relation to the tube axis.
**3.** Why are the outer rings set at $45^0$ to the axis of the tube?
**4.** Are the tube walls layered in a simple spiral wind? The alternative would be concentric tubes, referred to as the "Russian doll" mode.

The two types of hexagonal tiling associated with a triskelion were ruled out by the lack of six-fold, or three-fold symmetry in the X-ray images. The rectangular "brickwork" tilings with "T"-shaped triskelion vertices do not generate the $45^0$ rotation of the outer ring lobes relative to the central two-dimensional "ladder". We therefore consider a square tiling with a four-fold (tetraskelion) vertex (Figure 5) such as may exist when 1494Da polymer rods (Figure 6) are joined via silicon atoms (Figure 7A). Two possibilities for the vertex are illustrated in Figure 7. Each has a square pattern of iron atoms set at $45^0$ to the lattice rods, but only version 7A fits the diffraction data in respect to the inter-vertex spacing and the diffraction angle from the iron grouping. Version 7B is thought to be the most relevant one to the space-filling structure of section 3.

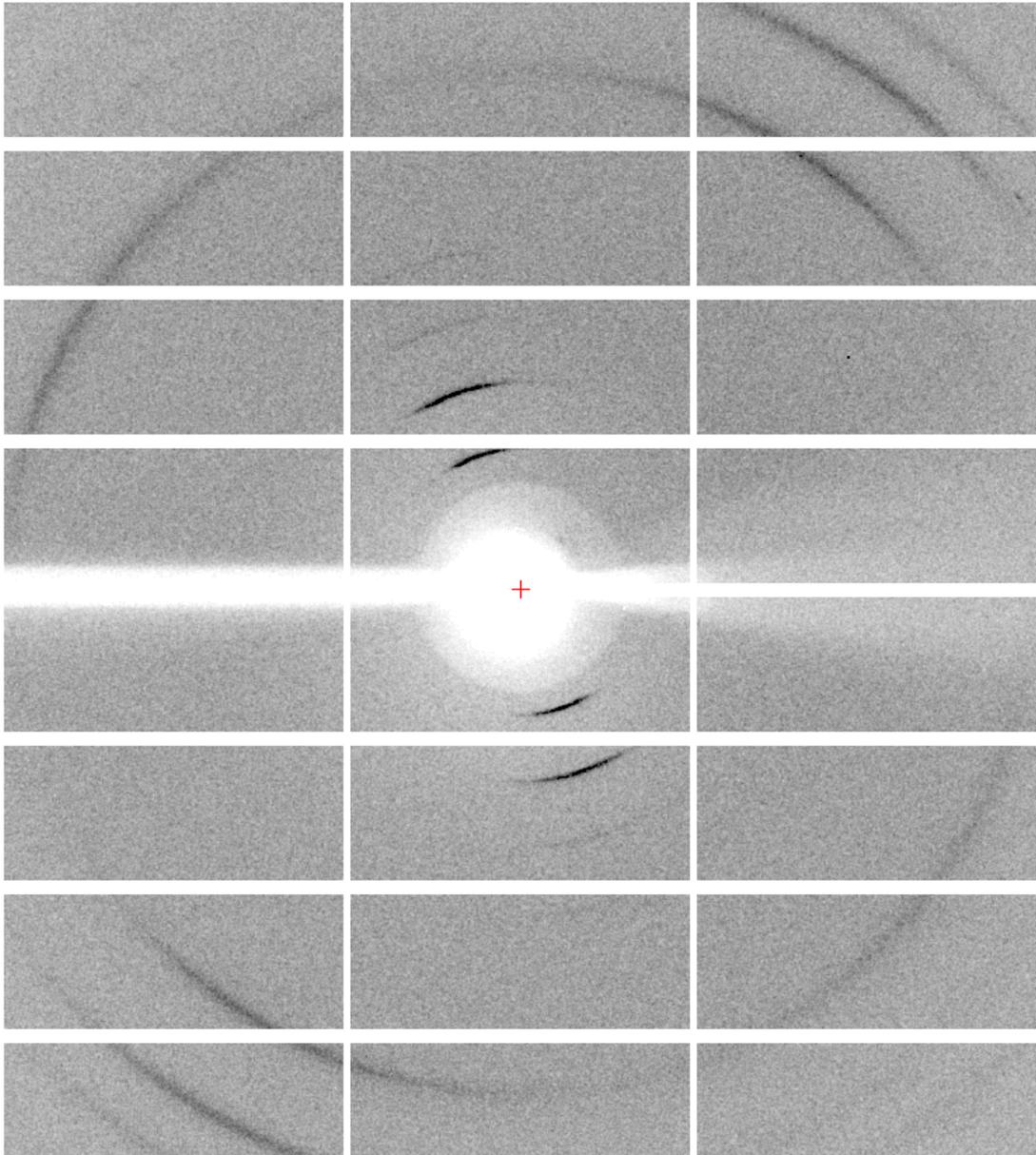

**Figure 4. X-ray room temperature (295K) diffraction pattern from Acfer 086 fiber crystals. Wavelength 0.979A, distance to detector 800mm. Innermost short arcs are nominally 24.19A, surrounded by features at 16.12A and faintly at 12.06A. First outer ring is at 6.31A.**

Square lattices were constructed using Spartan software [5,6] that includes MMFF energy minimization and Q-Chem functionality. Firstly, a single 1494Da molecule was generated (Figure 8A) in accord with the mass spectroscopy structure [1]. Each such rod was linked to three others by silicon atoms, with vertex details as shown in Figure 7A. Stacks of tetraskelia up to 3 layers deep were assembled and energy-minimized using MMFF, one example being shown in Figure 9. The separate 1494Da rods were found to be bound to each other along "edges" by hydrogen bonds with an energy of 71kJ/mol totaled along the whole length of each contact edge. Attempts to stack the 1494 Da rods in the thin, or "flat" direction did not generate hydrogen bonding, so this possibility was ruled out.

Measurements of the vertex-to-vertex distance on this simulated square lattice, after energy minimization via MMFF, gave a side $h$ = 48.38 ± 0.2 A (12 measurements). At each vertex in each layer there are 4 Fe atoms that dominate the X-ray scattering, consequently we test this calculated side $h$ against the X-ray data, with results shown in Table 1.

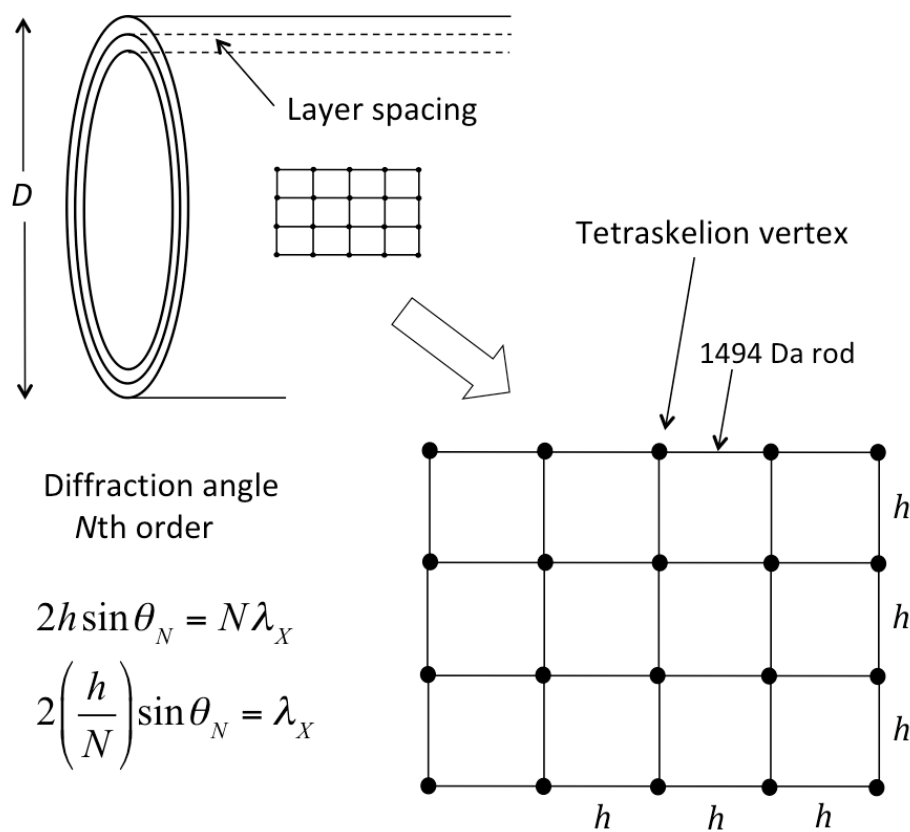

**Figure 5. Sketch of X-ray diffraction at wavelength $\lambda_X$ from a multiple-walled polymer amide nanotube based upon squares of side $h$ equal to the leg length of a tetraskelion 1494Da molecule. Repeat distance on circumference is $h$.**

**Table 1. Diffraction orders observed in central "ladder" with MMFF model for 1494 Da. rod polymer in squares of side *h*.**

| Diffraction order | *N*=2 | *N*=3 | *N*=4 | *N*=5 | Model *h* (A) |
|---|---|---|---|---|---|
| Observed (A) | 24.19 | 16.12 | 12.06 | 9.64 | |
| Fundamental (*N* x observed) | 48.38 | 48.36 | 48.24 | 48.2 | 48.38 ± 0.2 |

There is very good agreement between the modeled side of a square tiling of 1494 Da rods and all of the diffraction orders in the central motif. The observed central motif is a conserved "ladder" and not a "cross" because there is only one region of the tube wall that is always aligned to diffract in the direction perpendicular to the plane containing the tube axis.

On the supposition that the 1494Da polymer is the structural unit of square tiling, the strongest diffraction signals will come from the four iron atoms at each tetraskelion vertex (Figure 7A). The four-lobed diffraction rings at 6.31 and 5.29A originate from the $45^0$ planes with Fe atoms drawn in Figure 7A. These diffract to give relatively narrow rings because there is a depth of possibly several hundred layers in alignment through which these smaller square patterns extend. The calculated Fe spacing from MMFF modeling is 5.87 ± 0.11A (Figure 7A), which lies between the principal 6.31A and lesser 5.29A readings in diffraction. We conjecture that there are two possible iron locations, causing a split in this ring. The space of minor variations is too large to easily explore in such a complex situation.

The MMFF assembly of Figure 9 has a layer depth of 9.2A, but this spacing may a) be too sensitive to tube curvature to give prominent diffraction or b) be variable because there is not a radial constraint on the layers. When an additional lattice length $h$ is added to expand the circumference by one lattice unit the radius increase is $h/2\pi$ = 7.6A, which is not a perfect match to the estimated layer spacing of 9.2A, tending to rule out concentric tube ("Russian doll") packing.

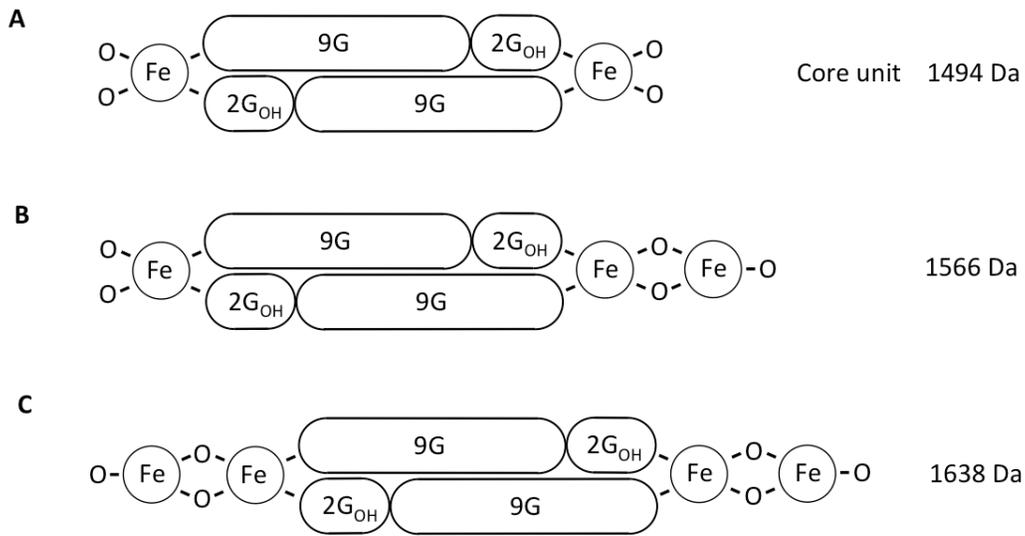

**Figure 6. Proposed structures for: A, the triskelion core unit at 1494Da [1]; B, the dominant 1566 Da entity in mass spectroscopy; C, the related dominant entity at 1638 Da, which is the proposed polymer rod in three-dimensional structures. In the models G = glycine and $G_{OH}$ = hydroxyglycine.**

In this set of X-ray diffraction images there is no sign of the connecting glycine polymers, which is not surprising in view of a) the much smaller diffraction of light elements relative to iron, and b) the open nature of the lattice, which does not constrain the polymer position.

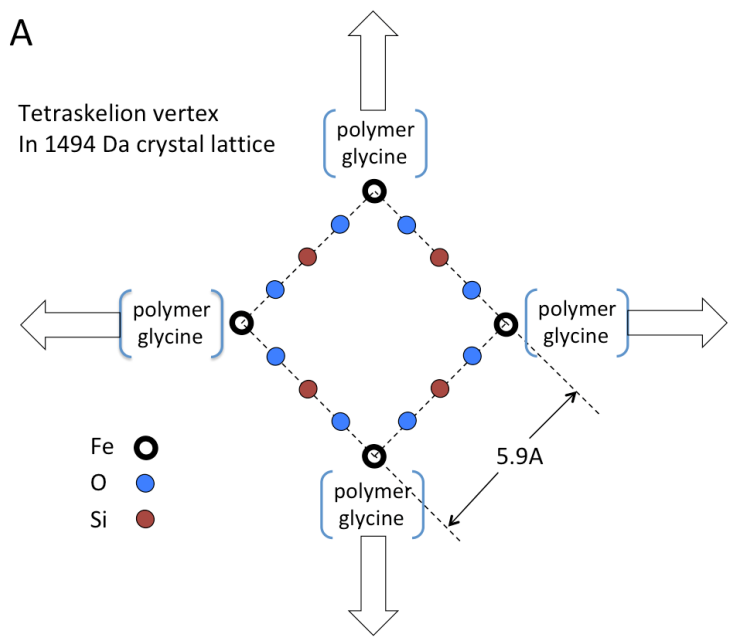

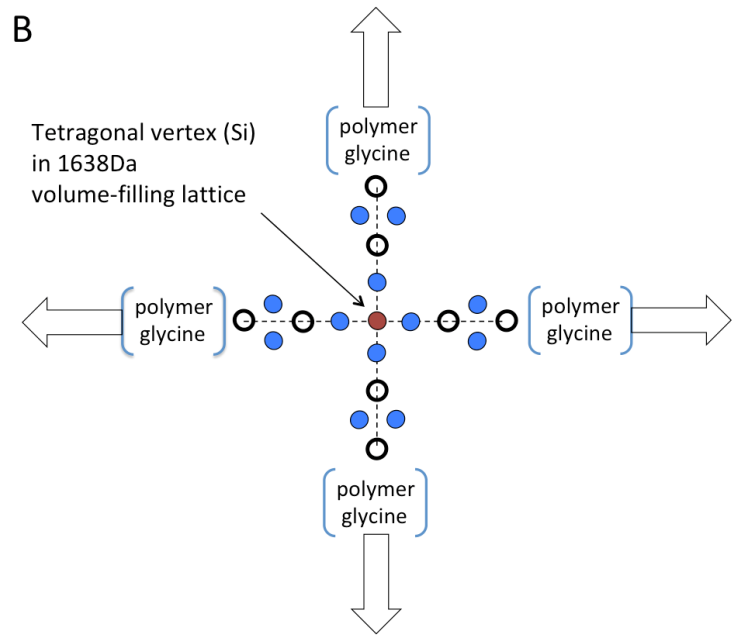

**Figure 7A Planar tetraskelion vertex in fiber crystal showing 45⁰ diffraction off iron atoms. Figure 7B Proposed tetragonal vertex in optimal space-filling structure.**

A

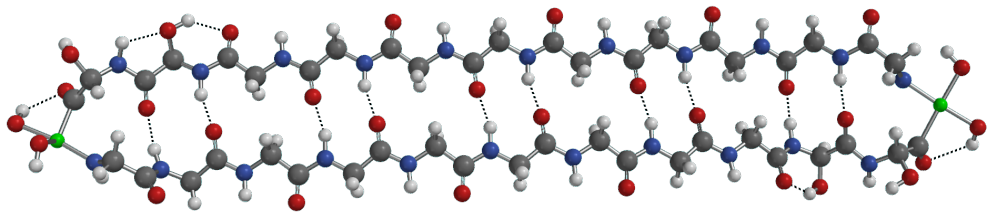

B

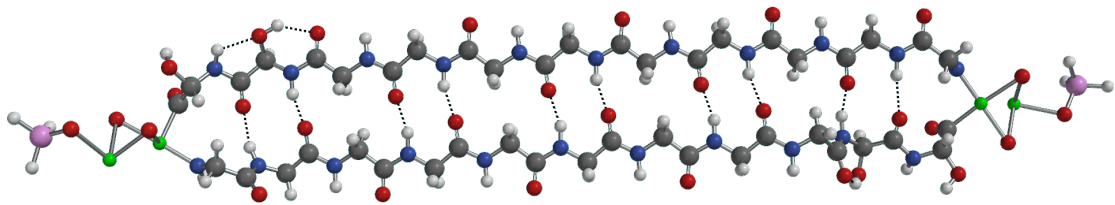

**Figure 8A. 1494Da rod built in Spartan/Q-Chem and subjected to MMFF at 298K. Figure 8B. 1638Da rod including tetragonal Si connector atoms. Ball & spoke format. Atom labels: Hydrogen white, carbon black, nitrogen blue, oxygen red, silicon pink and iron green.**

The present X-ray data indicates that the small crystals of Figure 3 are multiple-wall nanotubes composed of rolled-up sheets of a square lattice of 1494Da polymers linked at four-fold vertices which are a direct analog of the three-fold vertices proposed for the 4641Da polymer [1].

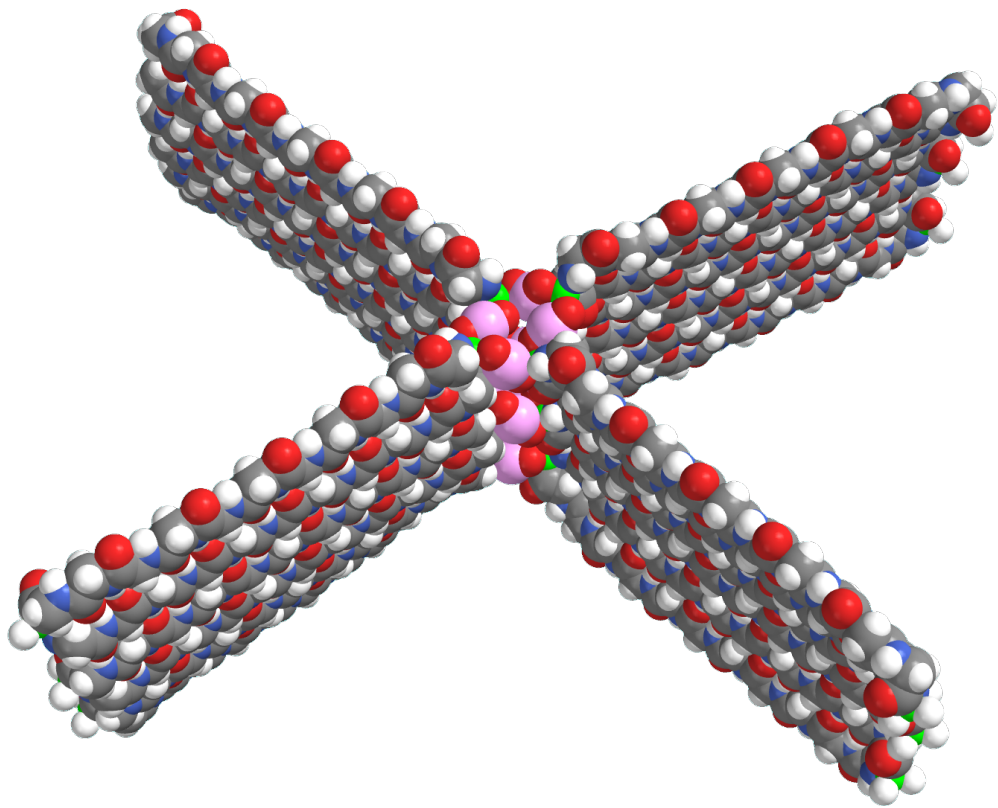

**Figure 9 Stacked 1494 Da tetraskelia in space-filling mode. Three layers hydrogen-bonded edge-to-edge. Hydrogen white, carbon black, nitrogen blue, oxygen red, silicon pink and iron green.**

**3. A volume-filling structure: calculation of the optimized specific volume**

Molecular clouds are regions within the high vacuum of space where there is low level content of hydrogen, helium and small molecules containing hydrogen, lithium, carbon, nitrogen and oxygen [7], extending to ethylene glycol and poly-aromatic hydrocarbons [8]. It is predicted that glycine, the simplest amino acid, should be able to form [9,10,11] but it has not been observed, possibly due to its polymerization which is exothermic [9] in the conditions of "warm, dense, molecular clouds". Also present are very small amounts of heavier elements such as magnesium, silicon and iron, providing a basis for the polymer amide observed [1,3], which is dominated by glycine and iron. In meteoritic material derived from asteroids we have observed a number of glycine polymers in relatively simple mass spectra up to m/z 4641 and dominant rod-like polymers around m/z 1600.

If, in the near vacuum of a pre-solar molecular cloud or of the proto-solar disc, there had been a path to the formation of these polymers, using incoming glycine, iron and water, then there would have been an advantage associated with presenting the largest possible cross-sectional area to intercept the most raw material. The polymer rod super-assembly with the greatest capability to span volume would then be enhanced via exponential competition above less optimal ones. This structure would naturally have very low density.

The observation (Figure 2) of phantom angled structures of total volume greater than the volume of meteoritic material initially added to a vial, was an indicator that the components of a very low-density network were present in the meteorite before solvent extraction. In each of the Folch extracts of KABA, Acfer 086 and Allende these components gravitated (upward) through the dense chloroform phase into the relatively thin interphase region, where they became sufficiently concentrated to re-assemble while floating as if in space. Isotopes at extraterrestrial levels [1,4] and the signature of a -C-C-N- amino acid polymer backbone in secondary ion mass spectrometry of the same Acfer 086 sample [12] were proof that these polymer rod subunits (around m/z 1600) pre-existed in the meteorite and could not have formed in the vial.

With a known, small set of dominant polymers to consider (identified by mass spectrometry) we propose here that the 1638Da polymer (observed at m/z 1639) forms the connecting rods of a largely empty space-filling lattice. The question we address is the likely nature of this lattice, given the materials available and the unique macro-scale morphology of angled shapes that has to be explained.

We will consider the creation of uniform repeating truly three-dimensional structures from a single type of rod, of length $h$ connected at either 4-fold or 6-fold vertices. The vertices are constrained to be of a single type in each of these cases, so that a single definite chemistry can apply throughout a structure. For example, the 4-fold vertex can be based upon a group IV atom such as carbon, silicon or germanium, with tetrahedral bond symmetry. Also, the connecting polymer rods will have end-for-end symmetry so as to form the lowest-entropy structure.

Just as the triskelion [1] in two dimensions can most efficiently create an enveloping surface of hexagons (and occasional pentagons) using a minimum of polymer rods, so in the third dimension we are led to consider flat hexagonally-tiled sheets containing rods of length $h$ connected perpendicularly by similar polymer rods that project "up" or "down" from neighboring vertices. One is immediately tempted to stretch this assembly vertically, so as to envelop more space with the same number of rods (Figure 10, parts A – E, high vertices labeled "H", low vertices labeled "L"). This puckers the hexagons so that their component rods make an angle $\alpha$ with the original plane (Figure 10A). Vertical stretching initially increases the volume of structure per polymer rod but the effect fades as the hexagon area decreases, its effective side length (viewed from above) reducing as $h\cos\alpha$ (Figure 10D). In terms of angle $\alpha$ the volume of a quasi-cell defined at its ends by two of these distorted hexagons is

$$\overline{V} = \frac{3\sqrt{3}}{2}h^3(1+\sin\alpha)\cos^2\alpha \qquad (2)$$

Setting $\frac{d\overline{V}}{d\alpha} = 0$ we find that a maximum in $\overline{V}$ exists at $\sin\alpha = 1/3$ i.e. $\alpha = 19.471^0$, when

$$\overline{V}_M = \frac{16\sqrt{3}h^3}{9} \qquad (3)$$

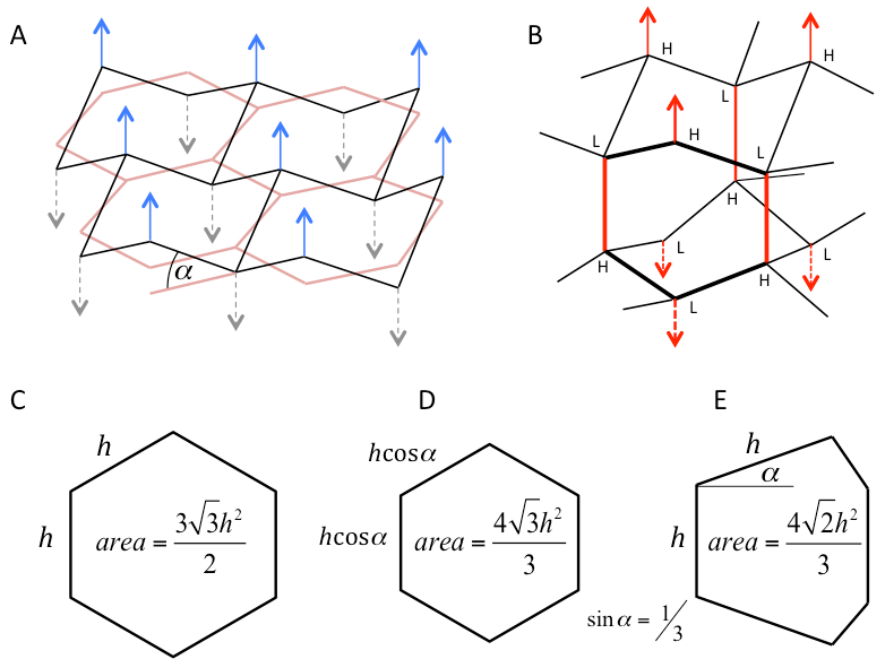

**Figure 10. A, Plane of hexagons (beige) distorted at alternate vertices (black). B, Interconnections between planes (red rods). C, Undistorted hexagons of side $h$. D, Projection of distorted hexagons of side $h\cos\alpha$, area applies for optimized $\sin\alpha = 1/3$. E, view along one of the 6 perpendicular channels, area applies for $\sin\alpha = 1/3$.**

The optimized volume for this stretched structure therefore occurs when the angle between one of the vertical connectors and any other rod at a vertex is the tetrahedral angle of $(90^0 + \alpha) = 109.471^0$. The structure with the optimized volume turns out to have exact tetrahedral symmetry at each vertex. By following this approach we have arrived at the 2H, or hexagonal diamond structure [13] that has three-fold rotation symmetry around the hexagon axis, and mirror symmetry in the plane centered between the quasi-planes of the hexagonal mesh. The 2H unit cell is smaller than those of 3C cubic diamond and higher polytypes such as 4H and 6H, so it is the most simple of all these structures.

In place of the carbon in a diamond structure we propose that Si, another group IV element with tetrahedral bonding symmetry, is the atom at the position of each vertex in our space-filling structure of polymer rods. Silicon is plentiful in these meteorites and is available as a free ion in aqueous solution. The triskelion linker atom was identified as silicon in mass spectrometry [1] making Si our leading candidate for the structural linker between polymer rods in three dimensions.

## 4. Properties

For the connecting polymer rods in 3D we invoke the 1638 Da polymer (seen at m/z 1639 and illustrated in Figures 6 and 8) because it is available in quantity (it and the related 1566 Da polymer are dominant in the mass spectrum [1] ) and it is end-for-end symmetrical, with an oxygen atom at each tip for bonding to Si. Effectively, a tetrahedral $SiO_4^{4-}$ anion becomes the central entity at a vertex (Figure 7B), with bonding to four surrounding Fe atoms. Each Fe atom is then linked to the pair of oxygen atoms at an end of a 1494 Da unit (Figure 6).

An MMFF simulation [5,6] (at 298K) gives the length of the 1638Da rods as $h$ = 4.806nm, measured from vertex to vertex, to the central Si atom positions (Fig.8B).

A measure of space-filling efficiency is the quotient $Q = \overline{V}$ /(number of rods per quasi-cell). Each quasi-cell contributes net 4 connecting rods because the six in the hexagon defining the end of a cell are each shared by a neighboring hexagon, making net 3 per cell and there is net one additional vertical connector per cell between the two hexagon ends. Consequently

$$Q = \frac{\overline{V_M}}{4} = \frac{4\sqrt{3}h^3}{9} \qquad (4)$$

From (4), using for the polymer rod the connector mass of 1638Da plus net ½ Si atom (14Da), we derive a lattice density of 32 mg cm$^{-3}$, when empty. In practice we observe the phantom angled shapes floating in the interphase region between water/methanol above and chloroform below so the structure that we see in our vials is permeated with a mixture of these solvents.

A related number of interest is the density of vertices. With one (distorted) hexagon per quasi-cell and each of its 6 vertices shared between three neighboring hexagons in the quasi-plane, there are net 2 vertices per quasi-cell. This can also be seen from the 4 rods per quasi-cell, each rod having 1/4 of a vertex at each of its ends, i.e. 1/2 of a vertex net per rod, or 2 vertices per quasi-cell. From the quasi-cell volume in (2) above, we find that the number $M$ of vertices per unit volume is given by

$$M = \frac{2}{V_M} = \frac{9}{8\sqrt{3}h^3} \qquad (5)$$

Using our calculated value of $h$ = 4.806nm, we find $M$ = 5.85 x $10^{18}$ cm$^{-3}$.

Turning to the simple cubic structure, the unit volume of $h^3$ uses only 3 rods because the 12 rods defining a cube are each shared with four neighboring cubes. For cubes we would need to devise a 6-fold chemical vertex with orthogonal axes. Comparing the efficiency of the above hexagonal diamond to the simple cubic structure as a creator of volume per unit number of polymer rods, the ratio is $4\sqrt{3}/3 = 2.309$ in favor of the former. The simple cubic alternative would be expected to produce copious $90^0$ angles, which are not observed.

## 5. From polymer to macroscopic realization: diffusion and symmetry

With the aid of a wire model it is possible to clearly see two types of open channels running through the hexagonal diamond 2H structure. The first channel type is exactly hexagonal in cross section and runs perpendicular to the quasi-planes. This first set of channels, parallel to the "hexagonal" axis, is intersected by 6-fold channels of a second type aligned perpendicularly to it and distributed around it at $60^0$ angular spacings. The latter channels are not exactly hexagonal in cross section, but rather have the cross section shown in Figure 10E, or the mirror image of 10E. They alternate at $60^0$ angles, yielding overall three-fold rotational symmetry. Altogether there are seven directions of wide-open channels within the structure that can be fast conduits for molecular diffusion. Other directions would necessitate zig-zag motion and have much slower diffusion rates. The hexagonal direction channels have a slightly larger cross section than the 6-fold perpendicular channels, in the ratio $\sqrt{3/2} = 1.225$. The faster degree of diffusion in the hexagonal axis direction may explain the generally linear trend of the structures. Elsewhere, the $60^0$ angles between transverse channels can explain the angled bends that are observed.

With this type of structure it is possible to understand how a concentration gradient would lead to the delivery of subunit rods in one or a few of the seven internal directions, leading to a surface growth of the structure that was not uniformly directed. Moreover, the angles between growth directions would often differ by $60^0$, as observed.

## 6. Discussion and Conclusions

Meteoritic polymers of glycine with iron based on a 1494Da core unit have been assigned rod-like character based upon analysis by mass spectrometry [1] and have been noted to assume a variety of characteristic macroscopic morphologies that beg explanation. Re-hydration of an extract from Acfer 086 gave X-ray diffracting crystals with a pattern characteristic of a multiple-walled nanotube. In the present nanotube crystal X-ray diffraction shows a square tiling with a lattice side of 4.836nm. The (MMFF, 298K) calculated length of the 1494Da rod identified in mass spectrometry, between tetraskelion vertices, is 4.838nm, in very good agreement with the diffraction measurement, also at 295 - 298K. The $45^0$ lobed diffraction rings at 6.31 and 5.29A are believed to be generated by the square pattern of iron atoms in each tetraskelion vertex. The deep stack of vertices generates a small ring thickness and the limited lateral iron spacing generates the lobes. However, the calculated pattern is at 5.87A, which may indicate that the iron location is split to generate the two rings on either side of this. The nanotube crystals do not necessarily represent structures pre-existing within the meteorite, but are the product of extraction, evaporation, sparing re-hydration and crystallization.

In a second structural arena, Folch solvent extracts of micron-scale meteoritic particles display floating small "phantom-like" tubules with frequent angled bends that suggest the influence of an underlying smaller-scale molecular structure [1]. In the same interphase layer where they are observed there is the highest mass spectrometry signal of glycine/ iron polymers of mass around 1600 Da with linear form [1]. Here we have proposed an explanation for the tubules in terms of a space-filling lattice of 1638 Da polymers linked by silicon atoms. We find that the simplest and most efficient space-filling lattice is the diamond 2H structure, with tetrahedral symmetry at the connecting vertices. By analogy with meteoritic triskelia [1] and because they have the requisite tetragonal symmetry we propose that the connecting vertices are silicon atoms. Here the lattice is permeated by solvent and floating in the layer that matches its density. The structure in free space has a calculated density of 32 mg cm$^{-3}$ and its regular internal passages have angle relationships of $90^0$ and $60^0$ that may guide molecular diffusion in the Folch solution to form the observed macroscopic angles. With such a low-density microscopic structure we can explain the apparently large volume of tubules relative to the meteoritic particles that released them.

It is possible that the accretion of planetary bodies in the proto-solar molecular cloud could initially have been driven by the development of such a space filling structure in near vacuum out of pre-existing glycine, iron and silicon [9]. Water, once trapped in such structures, would hydrogen bond preferentially around the more polar vertices, and eventually fill the structures, in the cool (100K) molecular cloud. With slight variations the polymers can make linear connecting tubes, surround air bubbles with vesicle walls, and fill space with a low-density lattice.

**Acknowledgements:** We wish to thank Guido Guidotti of Harvard for encouragement and advice for all of this extraterrestrial polymer research. The Acfer-086 meteorite sample was made available by Raquel Alonso Perez, curator of the Harvard Mineralogical and Geological museum. The KABA meteorite sample was made available at Harvard to J E M McG by the director, Professor, Dr. Béla Baráth and the deputy director, Dr. Teofil Kovács of the Museum of the Debreceni Reformatus Kollegium, Kalvin ter 16, H4026, Debrecen, Hungary. We thank Charles H. Langmuir and Zhongxing Chen for the use of the Hoffman clean room in the Harvard Earth and Planetary Science department. Shao-Liang Zheng is thanked for his initial X-ray analyses at the Harvard Department of Chemistry and Chemical Biology. Rachelle Gaudet and Jose Vellila of Harvard are also thanked for advice and assistance on crystallography techniques. The significant X-ray diffraction images were obtained on

# Structural Organization of Space Polymers

Julie E. M. McGeoch[1] and Malcolm W. McGeoch[2]

[1] Department of Molecular and Cellular Biology, Harvard University, 52 Oxford St., Cambridge MA 02138, USA.
[2] PLEX Corporation, 275 Martine St., Suite 100, Fall River, MA 02723, USA.

*Corresponding author. E-mail: mcgeoch@fas.harvard.edu

# Supplementary Information

### S1 METHODS
**Sample Source:** Meteorite sample Acfer-086 was made available by Raquel Alonso Perez, curator of the Harvard Mineralogical and Geological museum. The KABA meteorite sample was made available at Harvard to J E M McG by the director, Professor, Dr. Béla Baráth and the deputy director, Dr Teofil Kovács of the Museum of the Debreceni Reformatus Kollegium, Kalvin ter 16, H4026, Debrecen, Hungary.

**Meteorite Etch**: In a clean room extractor hood, at room temperature with high airflow, the samples were hand held with powder-free nitrile rubber gloves while being etched [1] to a total depth of 6mm with diamond burrs. The diamonds had been vacuum-brazed at high temperature onto the stainless-steel burr shafts to avoid the presence of glue of animal origin and organics in general. Etching on a fracture face (not an original exterior weathered face) was via slow steady rotation of a burr under light applied force via a miniature stepper motor that did not have motor brushes and did not contribute metal or lubricant contamination to the clean room. Two shapes of burr were used, the larger diameter type, in two stages, to create a pit of diameter 6mm and depth 6mm, and the smaller conical burr to etch a finely powdered sample, of approximately 1µm particles, from the bottom of the pit without contacting the sides. After each stage the pit contents were decanted by inversion and tapping the reverse side and a new burr was used that had been cleaned by ultrasonics in deionized distilled (DI) water followed by rinsing in DI water and air-drying in a clean room hood. The powder from the third etch was decanted with inversion and tapping into a glass vial and stored at -16C. Sample weights were in the range 2-8mg.

*From the start of this research 3D structures were noted in all methods of solvation of micron scale meteorite particles. However, the 3D structures were not the original aim of the research, that being centered on obtaining the correct solvation phase to produce enough polymer for MALDI mass spectrometry analysis [1] and extra-terrestrial isotope analysis [1,12]. Once the MALDI analysis had produced satisfactory mass data on the polymer we then concentrated on its 3D structural analysis. We considered the polymer would be in small amounts in the meteorites and would therefore be difficult to purify to obtain 3D crystals. Realizing it formed an extensive open lattice explained the many*

*3D forms observed on solvation. It then remained to find 3D forms that yielded meaningful information.*

*We first tested a variety of 3D forms in the Harvard Chemistry Department which indicated a mineral clino-enstatite was involved but there was evidence in those crystals of a 5nm periodicity consistent with the long glycine units in the polymer. We then decided to analyze all further crystals at APS, Argonne synchrotron.*

**Extraction of polymer amide via solvation of micron scale meteorite particles for 3D analysis:**

All extraction was performed at room temperatures between 17-23 Celsius.

Most solvation of the meteorite particles was by Folch extraction : a few mg of dry particles in chloroform/methanol/water, 3.3/2/1, in borosilicate 1ml and 2ml V-vials. This produced 2 phases, a heavy chloroform lower phase and a lighter polar phase. At the interphase a dark torus layer forms above the heavy chloroform bottom phase and beneath the top polar phase. Within hours the interphase torus layer consists of 3D triskelion and tetraskelion structures. Spherical 100micron diameter 3D structures also form from some Folch preparations at the interphase layer. These V-vials allow viewing of 3D structures while keeping the permanently closed vials upright with images taken at X10 magnification via an iPhone 11 pro max.

Other solvation was applied to dry particles on a glass slide viewed by light microscopy, involving all 3 Folch solvents (chloroform/methanol/water) and each separately.

Long time (many months) Folch extraction with very slow evaporation of the solvents followed by rehydration with water yielded the crystals that gave the best diffraction at the APS synchrotron (Fig. 4).

**X-ray analysis at APS beam line 31-ID-D.**

Yellow fiber crystals (Fig. S1.1) of polymer under light microscopy were attached to crystallography loops by dipping the loop into ethyl cyanoacrylate glue and then touching an end of the crystal with the glue loaded loop, ensuring the glue only contacted a distal part of the crystal. After the glue solvents had evaporated (4 hours at 18C) the crystals were kept for several days in their capped vials and examined daily to ensure they remained intact before shipping to APS synchrotron. Basically, these crystals are tough – they keep their 3D structure once formed and only degrade after many frames in the X-ray beam but can be shipped without refrigeration to a synchrotron via FEDEX if carefully wrapped to prevent undue vibration. After X-ray analysis the basic crystal changes to a white color from yellow due to bond damage. At the crystal base where no X-ray damage occurred the original yellow color persisted.

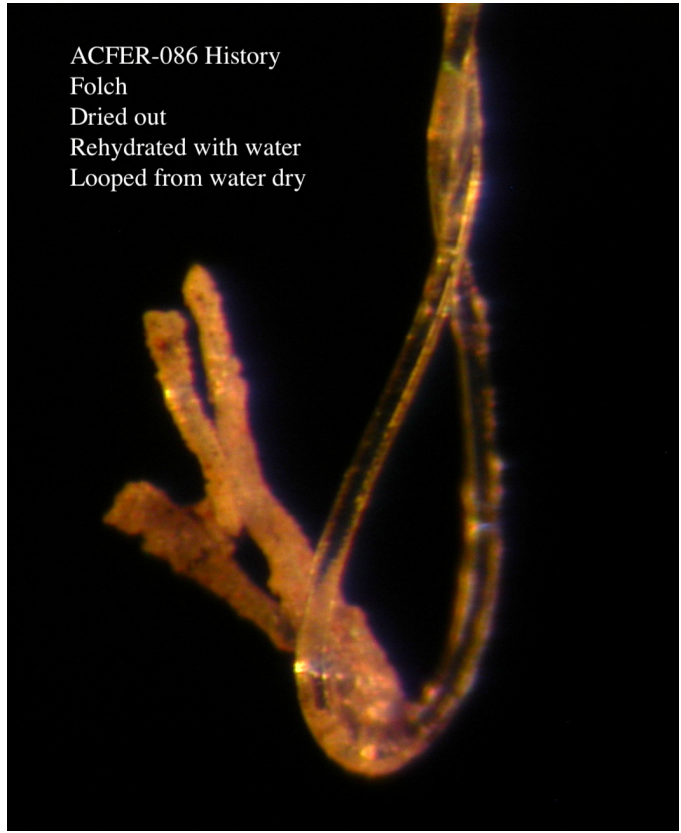

**Figure S1.1. Acfer-086 fiber crystal. The diameter of each of the 3 fiber crystals is approximately 50µm.**

The crystals were analyzed at 0.9793A. The raw APS X-ray data with detailed conditions is available on enquiry to mcgeoch@fas.harvard.edu.